\documentclass[12pt,a4paper]{article} 

\usepackage{feynmf,amsmath,amssymb,latexsym,epsfig,floatflt,subfigure}
\newcommand{\sst}{\scriptscriptstyle}
\newcommand{\st}{\scriptstyle}

\def\sst{\scriptscriptstyle}
\def\s {\small}

\newcommand{\be}{\begin{equation}}
\newcommand{\ee}{\end{equation}}
\def\bea{\begin{eqnarray}}
\def\eea{\end{eqnarray}}

\def\NPB#1#2#3{Nucl. Phys. {\bf B} {\bf#1} (#2) #3}
\def\PLB#1#2#3{Phys. Lett. {\bf B} {\bf#1} (#2) #3}
\def\PRD#1#2#3{Phys. Rev. {\bf D} {\bf#1} (#2) #3}
\def\PRL#1#2#3{Phys. Rev. Lett. {\bf#1} (#2) #3}

\def\NCA#1#2#3{Nuovo Cim. A {\bf#1} (#2) #3}
\def\ANPH#1#2#3{Ann. Phys. {\bf#1} (#2) #3}

\def\HEP#1{arXiv:hep-ph/#1}
\def\HEPEX#1{arXiv:hep-exp/#1}
\def\JMP#1#2#3{J. Math. Phys {\bf#1}(#2)#3}
\def\JHEP#1#2#3{JHEP {\bf#1} (#2) #3}
\begin{document} 
\thispagestyle{empty}
\begin{titlepage}

\title{\bf Spontaneous CP Violation in next to minimal renormalizable SUSY 
SO(10)}

\author{Yoav Achiman\footnote{e-mail:achiman@post.tau.ac.il}\\[0.5cm]                   
  School of Physics and Astronomy\\
  Tel Aviv University\\
  69978 Tel Aviv, Israel\\[1.5cm]}

\date{}
 
\maketitle
\setlength{\unitlength}{1cm}
\begin{picture}(5,1)(-12.5,-12)
\put(0,0){TAUP 2873/08}
\end{picture}
\parindent 0cm
\vspace{-2.0cm}
\begin{abstract}
\noindent
The minimal renormalizable SUSY SO(10) model (MSGUT), is a very compact and 
predictive
theory. It was very popular till one realized that it cannot account for the
masses of the neutrinos. The best cure to this problem is to add the
${\bf 120}$ Higgs representation, the ``{\em next to minimal}'' version, 
sometimes called ``new minimal susy GUT'' (NMSGUT).
To reduce the number of free parameters, it was suggested in 
recent papers to use only real parameters in the superpotential and induce 
CP violation via complex VEVs. This is what one usually calls {\em spontaneous 
CP violation}. The number of free parameters turned out, then, to be even 
smaller than in the original minimal model and good fits to all known masses 
and mixings were obtained.\\
Out of those papers,
only that of Aulakh and Garg discusses how CP is spontaneously violated.
Some heavy MSSM singlet VEVs generate a phase at high scale and CP violation 
is carried down to the CKM matrix by the mixing of the scalar MSSM doublets.
They study the model in great detail and give a large set of solutions.
As a proof of principle, two of the solutions are shown to induce realistic
phenomenological fits. It is not clear, however, how the right 
physical solution is obtained. The aim of this paper is to present a scenario 
how this can be done. I study the way 
solutions for spontaneous CP violation affect the scalar potential. The one
that gives the lowest minimum of the potential, in terms of a given set of
parameters, is the right physical one. In the way of doing so, I will prove 
that complex MSSM singlet VEVs lead actually to lower minima than the real (CP
conserving) ones. This proves that CP is spontaneously violated in this model.

\end{abstract}
\thispagestyle{empty} 
\end{titlepage}
 
\parindent 0pt 


\parindent=0pt

{\Large\bf  1\ \ \ \ \ Introduction}\\

SO(10) is the minimal GUT gauge group that involves naturally light massive
neutrinos\cite{so10} through the seesaw mechanism\cite{seesaw}.\\
Its supersymmetric version, SUSY SO(10), stabilizes the hierarchy and has 
R-Parity (matter parity) as a gauge symmetry. In the renormalizable case
R-Parity survives all symmetry breaking. Renormalizability requires high Higgs
representations, e.g. at least one ${\bf \overline {126}}$ Higgs.
 Therefore,
 the gauge coupling becomes ``strong'' (Landau pole) just above the GUT scale.
 Allowing for non-renomalizable contributions (suppressed by
$1/{ M_{\sst Planck}}$) one can get along with smaller representations.
(E.g. $ {\bf 16\times 16\approx {\overline {126}}}$ can play the role of     
${\bf\overline{ 126}}$ in SO(10)). On the other hand, renormalizable models 
require less ad hoc assumptions and fewer parameters than non-renormalizable 
ones. Their lightest SUSY particle is stable (LPS - a good dark matter 
candidate). Such models do not involve uncertain effects of gravitational
interactions. The fact that a Landau pole is not far from the GUT scale 
is not
a serious problem here, because one does not use the physics above it.\\
Quite a few papers discuss renormalizable SO(10) models
\cite{maryland}\cite{trieste}\cite{japan}\cite{wuppertal}\cite{AR}. 
The main attention was devoted recently to the minimal renormalizable 
SUSY SO(10) (MSSO(10)) \cite{minimal}. This
version is very compact and  predictive and several groups studied the model 
in great detail\cite{Senjan}\cite{Fukuyama}. One finds that
the requirement that SUSY remains unbroken at high energies  allows one to 
calculate the 
gauge symmetry breaking. This is related to the fermionic masses and mixing
through the fact that only two Higgs doublets remain light in the minimal
supersymmetric standard model (MSSM). The bi-doublet Higgs of the MSSM are
linear mixtures of all the original scalar doublets and the mixing parameters
depend on the way the gauge symmetry is broken.\\
Nice fits to the fermionic masses and mixing are obtained, except for the 
absolute masses of the neutrinos.  This is because gauge unification and other
reasons oblige the masses of the right handed (RH) neutrinos to lie not far 
from the GUT scale. This leads in the seesaw mechanism to too small masses of 
the neutrinos.
Recently suggested solutions involve adding the $D({\bf 120})$ Higgs
representation~\cite{120}, adding  type II seesaw\cite{II},
considering possible contribution from soft SUSY breaking terms\cite{soft} 
or adding warped extra dimensions\cite{extra}. \\ 
Adding  $D({\bf 120})$ Higgs is the only suggestion that is discussed in great
detail\cite{120}. The idea here is that when the $ H({\bf {10}})$ and  
$D({\bf 120})$ Higgs representations dominate the contributions to the 
fermionic masses, the Yukawa couplings of $\Sigma({\bf  {\overline{ 126}}})$
can be smaller and hence acquire smaller RH neutrino masses. This gives larger
neutrino masses and the right scale for leptogenesis.\\
However, the generic fits involve many parameters. To reduce the number of free
parameters, it was suggested recently to add the requirement that CP violation
should be generated spontaneously\cite{grimus}\cite{aulakh}. This means 
practically that all parameters in the
superpotential are real and CP violation is induced by complex VEVs. In this
way the number of free parameters is even smaller than in MSGUT.\\
Actually, spontaneous CP violation
 (SCPV) was already applied to SUSY SO(10) several years ago\cite{AR}. \\

Yet, the paper of Grimus and K\"uhb\"ock\cite{grimus} studies only generic fits to the fermionic masses and mixing using complex VEVs, without explaining how
the SCPV was generated. Aulakh and Garg\cite{aulakh}, on the other hand, 
discuss in detail the gauge symmetry breaking in the NMSGUT into the effective
MSSM (by fine tuning to keep a single pair of Higgs doublets light). They use
an analytic expression for all heavy MSSM singlet VEVs as solutions of a cubic
equation in a single variable ``$x$''. The Higgs fraction parameters which are
determined by the fine tuning condition are also functions of $x$. They prove 
then that the only way to have non trivial phases in the CKM matrix, for real
superpotential parameters, is for $x$ to be complex. Therefore, any complex
solution of the cubic equation leads to spontaneous CP violation. There
are, however,  many such solutions. As an example, two solutions are shown by 
the authors to give realistic fits to the fermionic masses and mixings.\\
Which one of those solutions is the correct physical one? 
Obviously, the solution that leads to the lowest minimum of the scalar
potential.\\
The aim of this paper is to suggest a scenario how this can be done. The idea
is to write the potential in terms of the scalar doublets and the heavy MSSM 
singlet VEVs. Then some VEVs are given a phase and one looks for the lowest 
minimum of the potential with respect to these phases . One can then fix $x$ 
and the corresponding Higgs fraction paremeters ``$\alpha_i^{u,d}$'', which 
dictate the CKM matrix. In the way of doing all this, I prove that complex
MSSM singlet VEVs lead to lower minima than the real (CP conserving) ones.
This proves that CP is spontaneously violated in the model.

The plan of the paper is as follows:
In Sector 2, I will give an introduction to SCPV. Sector 3 will 
present SUSY SO(10) and in particular the NMSGUT.
 Then in Sector 4 the SCPV in NMSGUT will be discussed in detail using an 
example. How CP violation is spontaneously generated
at the high scale and carried down to low energies will be summarized in 
Sector 5. The conclusions come in Sector 6.\\

\vspace{1.0cm}{\Large \bf 2  \ \ \ \ Spontaneous CP violation}\\

There are three manifestations of CP violation in Nature:\\

1) {\em Fermi scale CP violation} as is observed in the $K$ and $B$ 
decays\cite{KB}. This violation is induced predominantly by a complex mixing 
matrix of the quarks (CKM).\\

2) {\em The cosmological matter antimatter asymmetry (BAU)}
is an indication for high scale CP violation\cite{Sakharov}.
In particular, it's most popular explanation via leptogenesis\cite{Yanagida} 
requires CP breaking decays of the heavy right-handed (RH) neutrinos.\\

3) {\em The strong $CP$ problem } called also the QCD $\Theta$ 
problem\cite{QCD} lies in the non-observation of CP breaking in the strong
interactions while there is an observed CP violation in the interaction of
quarks.

\pagebreak
It is still not clear if there is one origin to those CP breaking 
manifestations.
What is the nature of the violation of CP? Is it intrinsic in terms of 
complex 
Yukawa couplings or due to spontaneous generation of phases in the Higgs 
VEVs ?\\

{\em Spontaneous violation of CP} \cite{T.D.Lee} is more difficult to realize,
but has advantages with respect to the intrinsic ones:\\

1) It is more elegant and involves less parameters. The intrinsic breaking 
becomes quite arbitrary in the framework of SUSY and GUT theories.\\

2) It solves the SUSY CP violation problem (too many potentially complex 
parameters) as all parameters are real.\\

3) It leads to the vanishing of $\Theta_{QCD}$ (but not 
ArgDetM) at the tree 
level. This can be used as a first step towards solving the CP problem 
by adding extra symmetries and exotic quarks
\cite{Nelson}\cite{Barr}\cite{branco}.\\

For good recent discussion of spontaneous CP violation, 
with many references, see Branco and Mohapatra\cite{BM}.\\

It is preferable to break CP at a high scale. This is what we need for
the BAU. Especially, if this is due to leptogenesis  i.e.
CP violating decays of heavy neutrinos, it is mandatory. This is also needed
to cure the domain wall problem~\cite{Zeldovich}.\\ 
Also, SCPV cannot take place in the standard model (SM) because of gauge 
invariance. Additional Higgs bosons must be considered and those lead 
generally to flavor changing neutral currents. The best way to avoid these is 
to make the additional scalars heavy\cite{BM}.\\

As a warm-up simple example for SCPV at the GUT scale let me present a possible
SCPV in the renormalizable non-SUSY SO(10).\\

SO(10) fermions are in three ${\bf 16}$ representations: $\Psi_i(\bf 16)$.
\be
{\bf 16}\times{\bf 16} = ({\bf 10} + {\bf 126})_S
+ {\bf 120}_{AS}\ .
\ee

Hence, only $H({\bf 10)},\ {\overline\Sigma}(\overline{\bf 126})$\ and\
$D({\bf 120})$ can contribute directly to Yukawa couplings and 
fermion masses. Additional Higgs representations are needed for the 
gauge symmetry breaking. \\
One and only one VEV \quad
${\bar\Delta} = <{\overline\Sigma} (1,1,0) >$\quad
can give a mass to the RH neutrinos via
\be
Y_\ell ^{ij} \nu_{\scriptscriptstyle{R}}^i {\overline\Delta}
 \nu_{\scriptscriptstyle{R}}^j 
\ee
and so induces the seesaw mechanism.
It breaks also B-L and SO(10) $\rightarrow$ 
SU(5).\\
To generate SCPV in conventional SO(10) one can use the fact that
${\overline\Sigma}({\overline {126}})$ is the only relevant complex Higgs 
representation. Its other special property is that $({\overline\Sigma})_
{\sst{S}}^{\sst{4}}$
is invariant in SO(10)\cite{HHR}. This allows for a SCPV at the high
scale, using the scalar potential~\cite{YA}:
\be
V = V_0 + \lambda_1(H)_{\sst{S}}^{\sst{2}} [({\overline\Sigma})_
{\sst{S}}^{\sst{2} }
+ ({\overline\Sigma}^*)_{\sst{S}}^{\sst{2}}] +
\lambda_2[({\overline\Sigma})_{\sst{S}}^{\sst{4}} + 
({\overline\Sigma}^*)_{\sst{S}}^{\sst{4}}]\ .
\ee
Inserting the Ansatz VEVs

\be
<H(1,2,-1/2)> = \frac{v}{\sqrt{\sst{2}}}
\ \ \ \ \  \overline\Delta = \frac{\sigma}{\sqrt{\sst{2}}}
{e^{i\alpha}}
\ee

in the neutral components, the phase dependent part of the scalar potential 
reads
\be
V(v,\sigma,\alpha) = A\cos (2\alpha) + B\cos (4\alpha)\ .
\ee

For $B$ positive and $|A| < 4B$ the absolute minimum of the 
potential requires
\be
\alpha = \frac{1}{2}\arccos\left( \frac{A}{4B}\right ). 
\ee
\nopagebreak 
This ensures  the spontaneous breaking of CP\cite{branco}.\\

It is not possible to realize the above scenario in renormalizable SUSY 
theories, as $\Phi^4$ cannot be generated from the superpotential in this 
case. A different approach is needed as will be presented later.\\

\vspace{2.0cm}

{\Large \bf 3\ \ \ \ The minimal renormalizable SUSY $SO(10)$  and next to 
minimal one}\\

Renormalizable SUSY SO(10) models were studied in many papers 
\cite{maryland}\cite{trieste}\cite{japan}\cite{wuppertal}\cite{AR}. 
In particular the so-called
{\em minimal renormalizable SUSY SO(10) model} (MSGUT)\cite{minimal} 
became very popular recently\cite{Senjan}\cite{Fukuyama}
due to its simplicity, predictability and automatic $R$-parity invariance
(i.e. a dark matter candidate).\\

It includes the following Higgs representations
\be
H({\bf 10}),\quad \Phi ({\bf 210}), \quad\Sigma ({\bf 126}) \oplus {\overline
\Sigma} ({\overline{\bf 126}})\ .
\ee
Both $\Sigma$ and $\overline{\Sigma}$ are required to avoid high scale SUSY 
breaking ($D$-flatness) and $\Phi ({\bf 210})$ is needed for the gauge 
breaking.
 \\

The properties of the model are dictated by the superpotential. This involves
all possible renormalizable products of the superfields

\be
W=M_\Phi \Phi^2 + \lambda_\Phi \Phi^3 + M_\Sigma \Sigma\overline\Sigma
+ \lambda_\Sigma \Phi\Sigma\overline\Sigma
+ M_{\scriptscriptstyle H} H^2 + \Phi H(\kappa\Sigma + \bar\kappa
\overline\Sigma) + 
\Psi_i(Y_{\scriptscriptstyle\bf 10}^{ij}H 
+ Y_{\overline{\scriptscriptstyle\bf 126}}^{ij} \overline\Sigma)\Psi_j
\ee
The symmetry breaking goes in two steps

\be
SUSY SO(10) \stackrel{strong\ gauge\ breaking}
{\longrightarrow} MSSM 
\stackrel{{\scriptscriptstyle SUSY}\ breaking}{\longrightarrow} SM
\ee

The $F$ and $D$-terms must vanish during the strong gauge 
breaking to avoid high scale SUSY breakdown ("$F$,$D$ 
flatness").\\

{\em $D$-flatness}:\quad
 only $\Sigma$, $\overline\Sigma$ are relevant, therefore
\be
|\Delta| = |\bar\Delta| .
\ee
The situation with\quad {\em $F$-flatness}\quad is more complicated.\\ 
The strong breaking is dictated by the VEVs that are SM singlets.\\ 
Those are, in the ${\scriptstyle SU_C(4)\times SU_L(2) \times SU_R(2})$  
notation :

$$
\phi_1 = <\Phi(1,1,1)> \ \  \phi_2 = <\Phi(15,1,1)> \ \  
\phi_3 = <\Phi(15,1,3)>
$$
$$
\Delta = <\Sigma(\overline{10},1,3)> \ \ \ \  \bar\Delta = 
<\bar\Sigma(10,1,3)> .
$$


The strong breaking superpotential in terms of those VEVs is then\footnote{
Using the notation of Ref. \cite{Fukuyama}.}
\be
\begin{array}{ccl}
W_H & = & M_\phi(\phi_1^2 + \phi_2^2 + \phi_3^2)
   + \frac{\lambda_\phi}{\sqrt {2}}(\frac{1}{9}\phi_1^3 + \frac{1}{2\sqrt{3}}
\phi_1\phi_2^2 + \frac{1}{3}\phi_2\phi_3^2)\\
[10pt]
&+& M_W\Delta\bar\Delta +\frac{ \lambda_\Sigma}{10}\Delta\bar\Delta
(\frac{1}{\sqrt {6}}\phi_1 +\frac{1}{\sqrt{2}}\phi_2 + \phi_3) .
\end{array}
\ee\\

$  \frac{\partial W_H}{\partial v_i} = 0$ \quad
gives a set of equations.
Their solutions dictate the details of the strong symmetry breaking.
\cite{trieste}\cite{Fukuyama}\\

One tunes the parameters such that the breaking 
$$
SUSY SO(10) \longrightarrow MSSM
$$
will be achieved\cite{Senjan}\cite{Fukuyama}.\\

The MSSM vacuum is fixed then by one parameter $x$, the solution of the cubic
equation\cite{trieste}:
\be
8x^2-15x^2+14x-3=-5\frac{\st \lambda_{\sst\Phi}\st M_{\sst \Sigma}}{\st \lambda_
{\sst \Sigma} \st M_{\sst \Phi}}(1-x)^2.
\ee

The high scale VEVs are then given as a function of $x$:\\
\be
\begin{array}{llr}
\Phi_1=-\frac{\st 2\sqrt{6} \st M_{\sst\Phi}}{\st \lambda_{\sst \Phi}}
\frac{x(1-5x^2)}{(1-x)^2},&\Phi_2=-\frac{2\sqrt{}18\st M_{\sst\Phi}}{\st 
\lambda_{\sst \Phi}}\frac{(1-2x-x^2)}{(1-x)},&\Phi_3=\frac{\st 12\st M_
{\sst\Phi}}{\st \lambda_{\sst \Phi}}x, \\ [10pt]
\Delta\bar\Delta=\frac{\st 240\st M_{\sst\Phi}^2}{\st \lambda_
{\sst \Phi}\lambda_{\sst\Sigma}} \frac{x(1-3x)(1+x^2)}{(1-x)^2}.& &
\end{array}
\ee

\pagebreak

SUSY is broken by the soft SUSY breaking terms.
The gauge MSSM breaking is induced by the VEVs
of the SM doublet $\phi^{u,d}(1,2,\pm 1/2)$ components of the Higgs 
representations.\\

The mass matrices of the higssinos and  Higgs scalars are:
\be
M_{ij}^u = \left[\frac{\partial^2W}{\partial\phi_i^u
\partial\overline{\phi_j^u}}\right]_{\phi_i=<\phi_i>}\ \ \ 
M_{ij}^d = \left[ \frac{\partial^2W}{\partial\phi_i^d
\partial\overline{\phi_j^d}}\right]_{\phi_i = <\phi_i>} .
\ee
The requirement 
\be
\det(M_{ij}^u) \approx 0 \ \ \ \det(M_{ij}^d)\approx 0
\ee
leaves only two light combinations of doublet components and those play the 
role of the bi-doublets \quad $h_u , h_d$ \quad of the MSSM. (This also is 
discussed in detail in the papers of~\cite{Senjan}\cite{Fukuyama}.) \\

However, as was explained in the introduction, the minimal model MSGUT 
cannot
account for the right neutrino masses. I  will study therefore  its minimal 
extension NMSGUT. One adds here the Higgs representation $D({\bf {120}})$
that couples antisymmetrically to the fermions. The fermionic mass matrices
can then be formally written as follows:
\begin{equation}
M^i=Y_{10}^i H +Y_{\overline {126}}^i \Sigma +Y_{120}^i D
\end{equation}{}
in terms of the Yukawa matrices, where $i=(u,d,e,\nu^{\sst D})$.\\

$D(\bf {120})$ does not involve MSSM singlets, hence it does not take part in 
the strong gauge breaking. I.e., the equations of the F,D-flatness are 
exactly as in the minimal model\cite{aulakh}. $D(\bf {120})$ contributes, however, new 
terms to the superpotential:\\

$W_D= \frac{ M_D}{\s 2}D^2 + {\s\lambda}_{\st 1}DH\Phi +
{\s \lambda}_{\st 2}DD\Phi + D\Phi({\s\lambda}\Sigma +
\bar{\s\lambda}\bar \Sigma )
+\Psi_i(Y_{\bf {\sst 120}}^{ij}D)\Psi_j$.\\

The MSSM relevant part of the superpotential includes the SM doublets:

\be
\begin{array}{ccccccc}
\phi^u&=&<\Phi(1,2,1/2)>&\ \ \ &\phi^d&=&<\Phi(1,2,-1/2)>\\[5pt]
H^u &=& <H(1,2,1/2)>&\ \ \ &H^d &=& <H(1,2,-1/2)>\\[5pt]
\Delta^u&=&<\Sigma(1,2,1/2)>& & \Delta^d&=&<\Sigma(1,2,-1/2)>\\[5pt]
\bar\Delta^u&=&<\bar\Sigma(1,2,1/2)>& &\bar\Delta^d & = &
<\bar\Sigma(1,2,-1/2)>\\[5pt]  
D_{\sst 1}^u&=&<D(1,2,1/2)>&\ \ \ &D_{\sst 1}^d&=&<D(1,2,-1/2)>\\[5pt]
D_{\sst 15}^u &=& <D(1,2,1/2)>&\ \ \ &D_{\sst 15}^d &=& <D(1,2,-1/2)>.
\
\end{array}
\ee\\
Note that $D({\bf 120})$ involves two kind of contributions under
${\st SU_{\st C}(4)\times SU_{\st L}(2) \times SU_{\st R}(2)}$ 

$$
D({\bf 120}): D^{u ,d}_{\sst 1} (1,2,2)\ \ \ \ 
 D^{u ,d}_{\sst 15}(15,2,2).
$$

\pagebreak

The scalar doublet mass matrix is now $6\times6$.\\

The VEVs of these doublets are linear combinations of its physical 
eigenvectors.\\ 
Using the fine tuning requirement (14), only the MSSM Higgs
doublets $h^u,h^d$ remain light\\

\be
\begin{array}{ccccccc}
\phi^{u,d}&=&\alpha_{\sst\phi}^{u,d}h^{u,d} &+&\ \ \ \ & heavy \ \   
(decoupled)\\[5pt]
H^{u,d}&=&\alpha_{\sst H}^{u,d}h^{u,d} &+&\ \ \ \ & heavy \ \   
(decoupled)\\[5pt]
\bar\Delta^{u,d}&=&\alpha_{\sst\bar\Delta}^{u,d}h^{u,d} &+&\ \ \ \ & heavy\ \ 
  (decoupled)\\[5pt]
\Delta^{u,d}&=&\alpha_{\sst\Delta}^{u,d}h^{u,d} &+&\ \ \ \ & heavy\ \ 
  (decoupled) .
\end{array}
\ee\\

Here $\alpha_i^{u,d}$ are the Higgs fractions, given  in explicit
complicate expressions in the paper of Aulakh and Garg\cite{aulakh}.
They play a crucial role in dictating the CKM matrix. The $\alpha_i^{u,d}$
are a function of $x$ and it was shown by Aulakh and Garg that for real values of $x$ the CKM matrix remain real. Hence, for real values of the superpotential
$x$ must be complex to have CP violation.\\

\vspace{2.0cm}
{\Large \bf 4 \ \ \ \ The spontaneous CP violation in NMSGUT}\\

Let us assume that all parameters of the superpotential as well as 
those of the soft SUSY breaking terms are real. CP will be violated 
spontaneously if certain VEVs generate a phase. In other words, the scalar 
potential will have a minimum with non-trivial phases.
 As was explained in
Sec. 2, we would like the phases to be generated for the heavy VEVs and
if possible also for  $\bar\Delta$, in order to have naturally leptogenesis. 
In terms of eq. (13) it is evident that complex MSSM singlet VEVs require
complex $x$. There are obviously a large set of such complex solutions of 
eq. (12). Two solutions are actually used in ref. \cite{aulakh} as a basis
for realistic generic fits for the fermionic masses and mixing. (One of them
corresponds to that of Grimus and K\"ub\"ock\cite{grimus}). The authors
emphasize themselves, however, ``that these fits are significant purely as 
proof of principle''. So, {\em which SCPVing solution is the right physical 
one?}. Clearly the solution that leads to the lowest minimum of the scalar
potential.\\
Let me present in the following a scenario how this can be done.\\ 

\pagebreak

The part of the effective superpotential (after D- and F-flatness are taken 
into account) that involves the coupling of the MSSM singlets to the doublets 
is as follows: \\

$W_ {\sst eff} =$
\be
\begin{array}{llll}
  \frac{\lambda_\Sigma}{\sst 10}
({\Phi}^u \Delta^d \bar\Delta + \Phi^d \bar\Delta^u\Delta)
&-& \frac{\kappa}{\sqrt{\sst 5}}\Phi^d H^u \Delta - 
\frac{\bar\kappa}{\sqrt{\sst 5}}
 {\Phi}^u H^d \bar\Delta     \\[7pt]
+\frac {\lambda_\Sigma}
{{\sst 15}\sqrt 2} \Phi_{\sst 2}(\bar\Delta^u\Delta^d +
{\bar\Delta}^d \Delta^u)& +& \frac{\kappa}{\sqrt{\sst 10}}
\Phi_{\sst 2}
(\Delta^d H^u - \Delta^u H^d)\\[7pt]
+\frac{\bar\kappa}{\sqrt{\sst 10}}\Phi_{\sst 2}
(\bar\Delta^d H^u - \bar\Delta^u H^d) &-& \frac{\kappa}{ 2
\sqrt{\sst 5}} \Phi_{\sst 3} (\Delta^d H^u + \Delta^u H^d) \\[7pt]
-\frac{\bar\kappa}{{ 2}\sqrt{\sst 5}}\Phi_{\sst 3} 
(\bar\Delta^d H^u
+ \bar\Delta^u H^d)&+&\frac{\lambda_\Phi}{6} 
\Phi^u \Phi^d (\frac{1}{\sqrt {\sst 2}}\Phi_{\sst 2}
+\frac{1}{2}\Phi_{\sst 3})\\[7pt]
-\frac{\lambda_{\sst 1}}{2} \Phi_{\sst 1}(D_{\sst 1}^u H^d +
D_{\sst 1}^d H^u) &-& \frac{\lambda_{\sst 1}}{2\sqrt{\sst 2}}
\Phi_{\sst 3}(D_{\sst {15}}^u H^d + D_{\sst{15}}^d H^u)\\[5pt]
+\frac{\lambda_{\sst }}{4\sqrt{\sst{30}}} \Phi_{\sst 3}
(D_{\sst 1}^u \Delta^d +
D_{\sst 1}^d \Delta^u) &+&
 \frac{\lambda_{\sst }}{6\sqrt{\sst{10} }}
\Phi_{\sst 3}(D_{\sst {15}}^u \Delta^d - 
D_{\sst{15}}^d \Delta^u)\\[7pt]
+\frac{\lambda_{\sst }}{4\sqrt{\sst{15}}} \Phi_{\sst 1}
(D_{\sst{15}}^u \Delta^d +
D_{\sst {15}}^d \Delta^u) &+& \frac{\bar\lambda_{\sst }}
{4\sqrt{\sst {30}}}
\Phi_{\sst 3}(D_{\sst {1}}^d \bar\Delta^u + 
D_{\sst{1}}^u \bar\Delta^d)\\[7pt]
+\frac{\bar\lambda_{\sst }}{6\sqrt{\sst{10}}} \Phi_{\sst 3}(D_{\sst {15}}^d
 \bar\Delta^u -
D_{\sst {15}}^u \bar\Delta^d) &+& \frac{\bar\lambda_{\sst }}{4\sqrt{\sst {15}}}
\Phi_{\sst 1}(D_{\sst {15}}^d \bar\Delta^u + D_{\sst{15}}^u \bar\Delta^d)\\[7pt]
+\frac{{\sqrt{\sst 2}}\lambda_{\sst 2}}{ 9} D_{\sst 15}^u D_{\sst 15}^d
\Phi_{\sst 2} &+& \frac{ \lambda_{\sst 2}}{6{\sqrt {\sst 3}}}(D_{\sst 1}^u
D_{\sst 15}^d+D_{\sst 15}^u D_{\sst 1}^d)\Phi_{\sst 3}\\[7pt]
-\frac{1}{2\sqrt{\sst{30}}} (\lambda_{\sst } D_{\sst 1}^u \Phi^d
 \Delta + \bar \lambda_{\sst }D_{\sst 1}^d \Phi^u \bar\Delta)&+
&\frac{\lambda_{\Sigma}}{30}\Phi_{\sst 3}(\bar\Delta^u\Delta^d-\bar\Delta^d
\Delta^u)\\[7pt]
-\frac{1}{2\sqrt{\sst {10}}} (\lambda_{\sst } D_{\sst {15}}^u
\Phi^d \Delta + \bar \lambda_{\sst } D_{\sst{15}}^d \Phi^u 
\bar\Delta).&&
\end{array}
\ee

Here the conventions of ref.\cite{Fukuyama} are used.\\

Note, that $W_{\st eff.}$ gives also the the mass matrix of the doublets\cite
{aulakh}\cite{Malinsky}. This mass matrix is fine tuned, see eq.(15).
The effect of this fine tuning to the MSSM will be taken into account, as in 
eq. (18), using the Higgs fractions. The MSSM singlet VEVs are not affected. 
$W_{\st eff.}$ is not the only source of the scalar potential,
other MSSM effective terms must be obviously added.
Let us, however, discuss first the part derived from 
$W_{\st eff}$. To prove SCPV, one must show that complex VEVs lead to a 
minimum of the scalar potential. In this case the Higgs fractions $\alpha_i^
{u,d}$ are also complex. Yet, the
different phases are correlated in view of their $x$ dependence. Hence, to
prove that CP is spontaneously violated, and to find out what are the physical
solutions of eq. (12), it is enough to show that two of the phases lead 
to a minimum.\\

The effective scalar potential of $W_{\st eff}$ is as follows
\\
\be
V_{\sst eff}(\lambda_{\st \Phi},\lambda_{\st \Sigma},\kappa, {\bar \kappa},
\lambda_i,
\lambda, {\bar \lambda}, V_i)=   \sum \left| \frac{\partial W_{\sst eff}}
{\partial V_i}\right|_{V_i}^2 .
\ee
 
Here $V_i$ stand vor the different VEVs and fields.  
This is a long and complicated expression. The corresponding derivatives can
be found in Appendix I. Note that instead of writing explicitly the derivative
with respect to e.g.\quad $\alpha_{\sst \phi}^u$,\quad we use the derivatives 
with respect to\quad $\Phi^u=\alpha_{\sst\Phi}^u h^u$\quad etc.\\

\pagebreak

Looking at the derivatives of $W_{\sst eff}$ in Appendix I, one sees that
the singlet VEVs appear always linearly. The most general phase dependence
of the singlet VEVs looks then as follows:
\be
\frac{\partial W_{\sst eff}}{\partial V_k}(\phi)=A^k + \sum B^k_j 
e^{i\st \phi_j}.
\ee 
Where  $A^k$ and $B^k$ are combinations of real coupling constants, real 
VEVs and the $\alpha_i$. We disregard here the phases of the 
$\alpha_i$, as they are any how correlated with the other phases. 


Hence,
\be
\begin{array}{ccl}
V_{\sst eff}&=&\sum_k \left| \frac{\partial W_{\sst eff}}
{\partial V_k}\right|_{V_k}^2=\sum_k \left|A^k + \sum_j B^k_j e^{i\st 
\phi_j}\right|^2 \\[10pt]
&=&\sum_{k,j} ({A^k}^2 + {B^k_j}^2 + 2A^k B^k_j \cos{\phi_j} +
 2\sum_{\ell \neq j}B^k_{\ell} B^k_j \cos{(\phi_{\ell}-\phi_j)}).
\end{array}
\ee\\
SCPV means here that some phases appear in the  minimum of the scalar potential
for a finite range of the parameters.\\
Let us look for special cases.\\ 
If only one VEV has a phase, the trivial solution\quad
$\phi=0,\pi$\quad results.\\
The simplest possibility is that two VEVs generate a phase. Let us 
stick to this possibility for simplicity (a generalization is straight 
forward).\\
Which phases should be involved?\\
$\bar\Delta$ is a most wishful candidate. Its phase will induce CP violation 
in the RH neutrinos decay, as is needed for leptogenesis (BAU). One cannot have
both  $\bar\Delta$ and $\Delta$ as candidates, no derivative involves both
of them (Appendix I).\\

So the simplest possibility is that $\bar\Delta$ and
one of the $\Phi_i$ will generate a phase $(\bar\delta, \phi)$ 
spontaneously.\\
One obtains in this case the following  generic scalar potential:

\be
V(\bar\delta, \phi)= S + R \cos{\bar\delta} +Q \cos{\phi} + 
T \cos{(\bar\delta - \phi)}.
\ee

Where,\\
$$
\begin{array}{ccccccc}
S&=&\sum(A_k^2+B_{\sst\bar\Delta_k}^2 + B_{\sst\Phi_k}^2),&\ \ \ \ 

&R&=&2\sum A_k B_{\sst\bar\Delta_k},\\[5pt] 
Q&=& 2\sum A_k B_{\sst\Phi_k},& \ \ \ \ \ \
&T&= &2\sum B_{\sst\bar\Delta_k}B_{\sst\Phi_k}.
\end{array}
$$\\
\nopagebreak
$S, Q, R$ and $T$  are combinations of coupling constants, VEVs and $\alpha_i$.
 They depend on the $\Phi_i$ one takes. They can be fixed in 
principle by the phenomenological fits and certain simplifying assumptions.\\

Now, to look for spontaneous generation of CP violation, we have to show that
there is a minimum of the scalar potential in a certain range of parameters,
with non-trivial values of the phases. (For the minimalization conditions
for two variables see Appendix II.)\\

\be
\begin{array}{ccccc}
\frac{\partial V}{\partial\bar\delta}&=&-R \sin{\bar\delta}-T \sin{(\bar\delta-
\phi)}&=&0\\ [10pt]
\frac{\partial V}{\partial\phi}&=&-Q \sin{\phi}-T \sin{(\bar\delta-
\phi)}&=&0
\end{array}
\ee\\

Solving the equations, one obtains
\be
\sin{\phi}=-\frac{R}{Q} \sin{\bar\delta}\ \ \ \ \ \ \ \ \     
cos{\bar\delta}=\frac{TQ}{2}
(\frac{1}{R^2}-\frac{1}{Q^2}-\frac{1}{T^2}).
\ee


The second derivatives are\\
\be
\begin{array}{ccc}
\frac{\partial^2 V}{\partial \bar\delta^2}&=& -R \cos{\bar\delta}-
T\cos{(\bar\delta-\phi)}\\[10pt]

\frac{\partial^2V}{\partial\phi^2}&=& -Q\cos{\phi}-T\cos{(\bar\delta-
\phi)}\\[10pt]

\frac{\partial^2V}{\partial\bar\delta\partial\phi}&=&T\cos{(\bar\delta-\phi)}.
\end{array}
\ee\\

The conditions for an extremum (Appendix II) require

\be
F\equiv(R\cos{\bar\delta} + T\cos{(\bar\delta-\phi)})
(Q\cos{\phi}+T\cos{(\bar\delta-\phi)})-T^2\cos^2{(\bar\delta-\phi)}>0
\ee\\
Using the above solutions (23), one obtains
$$
F=R^2\sin^2{\bar\delta}>0
$$
so that we have an extremum, {\em independent of the explicit expressions for 
$S,R,Q,T$}. \\
To have a minimum one needs also
$$
G\equiv-R\cos{\bar\delta}-T\cos{(\bar\delta-\phi)}>0.
$$
In terms of the solutions (25) it requires
$$
G=\frac{TR}{Q}>0.
$$

Hence, we have a non-trivial minimum for the range $TR/Q>0$.\\
This means that $\bar\Delta$ and one of the $\Phi_i$ generate spontaneously 
phases
at the high scale, for the range\quad  $\frac{\sst TR}{\sst Q}>{0}$.\quad CP 
is therefore violated at high energies. The values of the phases are given in 
eq. (25).\\

It is important to note that 
$$
\bar\delta=\phi=0 
$$
cannot be taken into account because it is a maximum. Therefore, {\em CP must 
be violated spontaneously in the NMSGUT model}.\\

The explicit expressions for $R,Q$ and $T$ depend on what $\Phi_i$ we 
choose, and  they are generally very complicated combinations of  
coupling constants, VEVs and the $\alpha_i$. It is useful to take 
$\bar\Delta$ and $\Phi_3$ as the corresponding VEVs.\\ 
Once phases of the complex singlets are known, one can, in principle, use 
$\Phi_3$ in eq. (13) to fix the value of $x$ and hence the Higgs fraction 
parameters $\alpha_i^{u,d}$ as well. Explicit expressions for the 
$\alpha_i$ as a complicate functions of $x$ are given in ref. \cite{aulakh}.
Those parameters dictate then the CP violating
CKM matrix as will be explained in the next section.\\



Now, we did not consider the other contributions to the scalar potential, and
in particular the effective MSSM Higgs potential and the soft SUSY breaking
terms. Those contributions, however, do not involve the singlet heavy VEVs . 
Hence, they can at most add a small contribution to the $A^k$
as they involve only low energy VEVs. The fact that the expression (23) mixes 
large values with small ones does not matter, because the phases are defined by 
ratios. The terms with heavy VEVs will decouple in the MSSM limit.
Note that, it is not surprising that high energy terms are involved in the 
scalar potential that dictates the phases. Also in ref. \cite{aulakh} the 
phases are generated at the high scale breaking.

\vspace{3.0cm}
{\Large \bf 5 \ \ \ \ \  CP violation at low energies}\\ 

We have seen that the scalar potential of the NMSGUT triggers SCPV at the 
high scale and the violation is carried down to low energies via the
$\alpha_i^{u,d}$.
At low energies, when the heavy fields decouple, the fine 
tuning condition (14) leads to the MSSM.
The effective MSSM superpotential, involves then the light Higgs fields, $h^u$
and $h^d$ with their Higgs fraction parameters $\alpha_i^{u,d}$, as is given 
by eq. (18). The Yukawa terms dictate the fermionic mass matrices of equ.(15)
as follows:

\be
\begin{array}{ccccc}
M_u&=&(\alpha^u_{\sst H}Y_{\st 10}+ \alpha^u_{\st {\sst\bar\Sigma}}Y_{\sst 
{\bar\Sigma}}
&+&(\alpha^u_{\sst D_{\sst 1}}+\alpha^u_{\sst D_{\sst {15}}})Y_{\sst 120})v^u
\\[10pt]
M_d&=&(\alpha^d_{\sst H}Y_{\st 10}+ \alpha^d_{\st {\sst\bar\Sigma}}Y_{\sst 
{\bar\Sigma}}
&+&(\alpha^d_{\sst D_{\sst 1}}+\alpha^d_{\sst D_{\sst {15}}})Y_{\sst 120})v^d
\\[10pt]

M_e&=&(\alpha^d_{\sst H}Y_{\st 10}-3 \alpha^d_{\st {\sst\bar\Sigma}}Y_{\sst 
{\bar\Sigma}}
&+&(\alpha^d_{\sst D_{\sst 1}}-3\alpha^e_{\sst D_{\sst {15}}})Y_{\sst 120})
v^d\\[10pt]
M_{\nu}^{Dirac}&=&(\alpha^u_{\sst H}Y_{\st 10}-3a^u_{\sst {\bar\Sigma}}
Y_{\sst {\bar\Sigma}}
&+&(\alpha^u_{\sst D_{\sst 1}}-3\alpha^u_{\sst D_{\sst {15}}})Y_{\sst 120})v^u.
\end{array}
\ee
\\
Where $Y_i$ are the corresponding Yukawa matrices and  $<h^u>=v^u$,
$<h^d>=v^d$ .\\

Complex $\alpha_i^{u,d}$ induce complex mass matrices. Hence, we obtain 
a CP violating CKM matrix.

\pagebreak           
\vspace{-1.0cm}
{\Large \bf 6 \ \ \ \ Conclusions}\\

Spontaneous CP violation has many advantages on the intrinsic breaking and is 
more natural. 
Nevertheless, SCPV has been rarely used in GUTs.\\
Recently two groups applied spontaneous CP violation in the renormalizable 
NMSGUT to reduce the number of free parameters\footnote{See also ref. \cite
{AR}}.\\
One of those paper by Aulakh and Garg\cite{aulakh} showes that SCPV is 
actually posible in NMSGUT. They proved that there are complex solutions to
the GUT scale cubic equation for $x$, and those lead to CP violation in the CKM matrix, via the the complex Higgs fractions $\alpha_i^{u,d}$. It is not clear,
however, what is the  right physical solution. \\
I have proven that CP is really violated in NMSGUT, by showing that the 
minimum of the scalar potential violates CP, in terms of MSSM singlet VEVs
with very specific phases. The phases that minimize the scalar potentioal can 
be used therefore to dictate the physical $x$. 
The corresponding Higgs fractions alows one then to get the physical complex
CKM matrix. One of the complex VEVs is that of the $\overline{\bf 126}$ ,
hence the needed high scale CPV for leptogenesis is also suppressed.  Also, the
spontaneous breaking is at the high scale, FCNCs and domain walls 
are avoided. The minimalization of the scalar potential completes 
therefore the programm of Aulakh and Garg.\\

{\large \ \ \ \ \ \ {\bf Note added }}\\

After the manuscript was finished, I learned about a preprint by Malinsk\'y
\cite{Malinsky}. He also discusses the Higgs sector of the NMSGUT, but without
restricting the parameters of the superpotential. He calculates the mass 
matrices in a way similar to Aulakh and Garg\cite{aulakh}, and finds 
explicitly the corresponding Higgs fractions. Malinsk\'y's results involve 
many free parameters as SCPV is not assumed. On top of that, because he uses
different methods and phase conventions it is difficult to compare his results
with those of Aulakh and Garg for the special case of SCPV. 
\pagebreak

{\Large Appendix I :The derivatives of the superpotential}\\

$\frac{\partial}{\partial\Delta} = $\\[5pt]
$\frac{\lambda_\Sigma}{10} \Phi^{d}{\bar\Delta^u}-\frac{\kappa}{\sqrt 5}
\Phi^d H^u - \frac{\lambda}{2\sqrt{\sst 30}} (D_{\sst 1}^u \Phi^d 
+ D_{\sst 15}^u \Phi^d) $ \\[5pt]

$\frac{\partial}{\partial\bar\Delta} = $\\[5pt]
$\frac{\lambda_\Sigma}{10} \Phi^{u}{\Delta}^d- \frac{\kappa}{\sqrt {\sst 5}}
\Phi^u H^d - \frac{\bar\lambda}{2\sqrt{\sst 30}} (D_{\sst 1}^d \Phi^u +
 D_{\sst 15}^d \Phi^u)$ \\[5pt]

$\frac{\partial}{\partial\Delta^d} = $\\[5pt]
$\frac{\lambda_\Sigma}{10}\Phi^u \bar\Delta  
+ \frac{\lambda_\Sigma}{15\sqrt{\sst  2}}\Phi_{\sst 2}\bar\Delta^u
+ \frac{\lambda_\Sigma}{30}\Phi_{\sst 3} \bar\Delta^u
+ \frac{\kappa}{\sqrt{\sst 10}} \Phi_2 H^u  - 
\frac{\kappa}{2\sqrt{\sst 5}}\Phi_{\sst 3} H^u\\[5pt]
+ \frac{\lambda}{4\sqrt{\sst 30}} \Phi_{\sst 3} D_{\sst 1}^u +
 \frac{\lambda}{6\sqrt{\sst 10}}\Phi_{\sst 3} 
D_{\sst 15}^u + \frac{\lambda}{4\sqrt{\sst 15}}\Phi_{\sst1} D_{\sst 15}^u
$\\[5pt]

$\frac{\partial}{\partial\Delta^u} = $\\[5pt]
$\frac{\lambda_\Sigma}{15\sqrt 2}\Phi_{\sst 2}\bar\Delta^d
-\frac{\lambda_\Sigma}{30}\Phi_{\sst 3} \bar\Delta^d
- \frac{\kappa}{\sqrt{\sst 10}} \Phi_{\sst 2} H^d -
 \frac{\kappa}{2\sqrt{\sst 5}}\Phi_{\sst 3} H^d
+ \frac{\lambda}{4\sqrt{\sst 30}} \Phi_{\sst 3} D_{\sst 1}^d\\[5pt]
 - \frac{\lambda}{6\sqrt{\sst10}}\Phi_{\sst 3} 
D_{\sst15}^d + \frac{\lambda}{4\sqrt{\sst 15}}\Phi_{\sst 1} D_{\sst 15}^d
$\\[5pt]

$\frac{\partial}{\partial\bar\Delta^d} = $\\[5pt]
$ \frac{\lambda_\Sigma}{15\sqrt{\sst 2}}\Phi_{\sst 2}\Delta^u
-\frac{\lambda_\Sigma}{30}\Phi_{\sst 3}\Delta^u
+ \frac{\bar\kappa}{\sqrt{\sst 10}} \Phi_{\sst 2} H^u  
+ \frac{\bar\lambda}{4\sqrt{\sst 30}} \Phi_{\sst 3} D_{\sst 1}^u  
-\frac{\bar\lambda}{6\sqrt{\sst 10}}\Phi_{\sst 3} 
D_{\sst 15}^u\\[5pt]
 + \frac{\lambda}{4\sqrt{\sst 15}}\Phi_{\sst 1} D_{\sst 15}^u$\\[5pt]

$\frac{\partial}{\partial\bar\Delta^u} = $\\[5pt]
$ \frac{\lambda_\Sigma}{15\sqrt 2}\Phi_2\Delta^d
+ \frac{\lambda_\Sigma}{30}\Phi_3\Delta^d
- \frac{\bar\kappa}{\sqrt{10}} \Phi_2 H^d - \frac{\bar\kappa}{2\sqrt{5}}\Phi_3 H^d
+ \frac{\bar\lambda}{4\sqrt{30}} \Phi_3 D_1^d\\[5pt] 
- \frac{\bar\lambda}{6\sqrt{10}}\Phi_3 
D_{15}^d + \frac{\bar\lambda}{4\sqrt{15}}\Phi_1 D_{15}^d
$\\[5pt]

$\frac{\partial}{\partial H^u} = $\\[5pt]
$-\frac{\kappa}{\sqrt 5}\Phi^d\Delta+\frac{\kappa}{\sqrt{10}} \Phi_2 \Delta^d
+\frac{\bar\kappa}{\sqrt10}\Phi_2\bar\Delta^d-\frac{\kappa}{2\sqrt5}\Phi_3\Delta^d
-\frac{\bar\kappa}{2\sqrt5}\Phi_3\bar\Delta^d\\[5pt]
-\frac{\lambda_1}{2}\Phi_1 D_1^d-
\frac{\lambda_1}{2\sqrt 2}\Phi_3 D_{15}^d$\\[5pt]

$\frac{\partial}{\partial H^d} = $\\[5pt]
$-\frac{\bar\kappa}{\sqrt 5}\Phi^u\bar\Delta-\frac{\kappa}{\sqrt{10}} \Phi_2 \Delta^u
-\frac{\bar\kappa}{\sqrt10}\Phi_2\bar\Delta^u-\frac{\kappa}{2\sqrt5}\Phi_3\Delta^u
-\frac{\bar\kappa}{2\sqrt5}\Phi_3\bar\Delta^u\\[5pt]
-\frac{\lambda_1}{2}\Phi_1 D_1^u
-\frac{\lambda_1}{2\sqrt 2}\Phi_3 D_{15}^u$\\[5pt]

$\frac{\partial}{\partial\Phi_{\sst 1}} = $\\[5pt]
$-\frac{\lambda_{\sst 1}}{2}(D_{\sst 1}^u H^d+ D_{\sst 1}^d H^u)
+ \frac{\lambda}{4\sqrt{\sst 15}}
(D_{\sst 15}^u\Delta^d + D_{\sst 15}^d\Delta^u) 
+ \frac{\bar\lambda}{4\sqrt{\sst 15}}
(D_{\sst 15}^d\bar\Delta^u + D_{\sst 15}^u\bar\Delta^d)$\\

\pagebreak
\vspace{-3.0cm}
$\frac{\partial}{\partial\Phi_{\sst 2}} = $\\[5pt]
$\frac{\sqrt{\sst 2}\lambda_\Sigma}{15}(\bar\Delta^u\Delta^d + \bar\Delta^d\Delta^u)
+ \frac{\kappa}{\sqrt{\sst 10}}(\Delta^dH^u - \Delta^uH^d)\\[3pt]
+\frac{\bar\kappa}{\sqrt{\sst 10}}(\bar\Delta^dH^u -\bar\Delta^uH^d) 
+\frac{\lambda_\phi}{6\sqrt{\sst 2}}\Phi^u\Phi^d +
 \frac{\sqrt{\sst 2}\lambda_{\sst 2}}{9}D_{\sst 15}^uD_{\sst 15}^d $\\[3pt]

$\frac{\partial}{\partial\Phi_{\sst 3}} = $\\[3pt]
$-\frac{\kappa}{2\sqrt{\sst 5}}(\bar\Delta^dH^u+ \Delta^uH^d) 
-\frac{\bar\kappa}{2\sqrt{\sst 5}}(\Delta^dH^u+ \bar\Delta^uH^d) \\[3pt]
-\frac{\lambda_{\sst 1}}{2\sqrt{\sst 2}}(D_{\sst 15}^uH^d + D_{\sst 15}^dH^u)
+\frac{\bar\lambda}{4\sqrt{\sst 30}}(D_{\sst 1}^d\bar\Delta^u 
+ D_{\sst 1}^u\bar\Delta^d)\\[3pt]
+\frac{\lambda}{6\sqrt{\sst 10}}(D_{\sst 15}^u\Delta^d - D_{\sst 15}^d\Delta^u)
+\frac{\bar\lambda}{6\sqrt{\sst 10}}(D_{\sst 15}^d\bar\Delta^u 
- D_{\sst 15}^u\bar\Delta^d)\\[3pt]
+\frac{\lambda_{\sst 2}}{6\sqrt{\sst 3}}(D_{\sst 1}^uD_{\sst 15}^d
 + D_{\sst 15}^uD_{\sst 1}^d)
+\frac{\lambda_\Sigma}{30}(\bar\Delta^u\Delta^d - \bar\Delta^d\Delta^u)$\\[3pt]

$\frac{\partial}{\partial\Phi^d} = $\\[3pt]
$-\frac{\kappa}{\sqrt{\sst 5}}H^u\Delta + \frac{\lambda_\Sigma}{10}\bar\Delta^u\Delta 
+\frac{\lambda_\phi}{6}\Phi^u(\frac{1}{\sqrt{\sst 2}}\Phi_{\sst 2}
+\frac{1}{2}\Phi_{\sst 3} )\\[5pt]
-\frac{\lambda}{2\sqrt{\sst 30}}(D_{\sst 1}^u \Delta + D_{\sst 15}^u\Delta)$\\[3pt]

$\frac{\partial}{\partial\Phi^u} = $\\[3pt]
$\frac{\lambda_\Sigma}{10}\Delta^d\bar\Delta 
- \frac{\bar\kappa}{\sqrt{\sst 5}}H^d\bar\Delta
+\frac{\lambda_\Phi}{6}\Phi^d((\frac{1}{\sqrt{\sst 2}}\Phi_{\sst 2}
+\frac{1}{2}\Phi_{\sst 3} )\\[3pt]
-\frac{\bar\lambda}{2\sqrt{\sst 30}}(D_{\sst 1}^d + D_{\sst 15}^d)\bar\Delta $\\[3pt]

$\frac{\partial}{\partial D_{\sst 1}^d} = $\\[3pt]
$- \frac{\lambda_{\sst 1}}{2}\Phi_{\sst 1}H^u 
+ \frac{\lambda}{4\sqrt{\sst 30}} \Phi_{\sst 3}\Delta^u
+ \frac{\bar\lambda}{4\sqrt{\sst 30}}\Phi_{\sst 3}{\bar\Delta^u} 
+ \frac{\lambda_{\sst 2}}{6\sqrt{\sst 3}}
\Phi_3 D_{\sst 15}^u \\[3pt] 
- \frac{\bar\lambda}{2\sqrt{\sst 30}}\Phi^u \bar\Delta $\\[3pt]

$\frac{\partial}{\partial D_{\sst 1}^u} = $\\[5pt]
$- \frac{\lambda}{4\sqrt{\sst 30}}\Phi_{\sst 3} \Delta^d 
- \frac{\lambda_{\sst 1}}{2}\Phi_{\sst 1}H^d
+ \frac{\bar\lambda}{4\sqrt{\sst 30}}\Phi_{\sst 3}{\bar\Delta^d}
 + \frac{\lambda_{\sst 2}}{6\sqrt{\sst 3}}
\Phi_{\sst 3} D_{\sst15}^d \\[5pt] 
- \frac{\bar\lambda}{2\sqrt{\sst 30}}\Phi^d \Delta$\\[5pt]

$\frac{\partial}{\partial D_{\sst 15}^d} = $\\[5pt]
$-\frac{\lambda_{\sst 1}}{2\sqrt{\sst 2}} \Phi_{\sst 3} H^u 
- \frac{\lambda}{6\sqrt{\sst 10}}\Phi_{\sst 3}\Delta^u
+ \frac{\lambda}{4\sqrt{\sst 15}}\Phi_{\sst 1}\Delta^u 
+ \frac{\bar\lambda}{6\sqrt{\sst 10}}\Phi_{\sst 3}
\bar\Delta^u \\[5pt]
+ \frac{\bar\lambda}{4\sqrt{\sst15}}\Phi_{\sst 1} \bar\Delta^u + 
\frac{\sqrt{\sst 2}\lambda_{\sst 2}}{9}\Phi_{\sst 2}D_{\sst 15}^u 
+ \frac{\lambda_{\sst 2}}{6\sqrt{\sst 3}}\Phi_{\sst 3} D_{\sst 1}^u
- \frac{\bar\lambda}{2\sqrt{\sst 10}}\Phi^u\bar\Delta$\\[5pt]

\nopagebreak
$\frac{\partial}{\partial D_{\sst 15}^u} = $\\[5pt]
$-\frac{\lambda_{\sst 1}}{2\sqrt{\sst 2}} \Phi_{\sst 3} H^d 
+ \frac{\lambda}{6\sqrt{10}}\Phi_3\Delta^d
+ \frac{\lambda}{4\sqrt{\sst 15}}\Phi_{\sst 1}\Delta^d 
- \frac{\bar\lambda}{6\sqrt{\sst 10}}\Phi_{\sst 3}
\bar\Delta^d\\[3pt] \nopagebreak
+ \frac{\bar\lambda}{4\sqrt{\sst 15}}\Phi_{\sst 1} \bar\Delta^d + 
\frac{\sqrt{\sst2}\lambda_{\sst 2}}{9}\Phi_{\sst 2} D_{\sst 15}^d 
+ \frac{\lambda_{\sst 2}}{6\sqrt{\sst 3}}\Phi_{\sst 3} D_{\sst 1}^d
- \frac{\lambda}{2\sqrt{\sst 10}}\Phi^d\Delta$.

\pagebreak

{\Large Appendix II: Minimum of a function with two variables}\\

Conditions for an extremum:
$$
a)\frac{\partial{\cal F}(x_0,y_0)}{\partial x}=
\frac{\partial{\cal F}(x_0,y_0)}{\partial y}=0
$$
$$
b)(\frac{\partial^2}{\partial x^2}{\cal F}(x_0,y_0))
(\frac{\partial^2}{\partial y^2}{\cal F}(x_0,y_0))-
(\frac{\partial^2}{\partial x\partial y}{\cal F}(x_0,y_0))^2 >0.
$$

${\cal F}(x_0,y_0)$ is a minimum if on top of a) and b)
$$
(\frac{\partial^2}{\partial x^2}{\cal F}(x_0,y_0))>0.
$$

\end{document}